\documentclass[prd,showpacs,preprint,nofootinbib]{revtex4-1}

\usepackage{amsmath,amssymb}
\usepackage[colorlinks=true,linkcolor=blue]{hyperref}

\newcommand{\nn}{\nonumber}
\def\/{\over}
\newcommand{\bra}[1]{\langle#1|}
\newcommand{\ket}[1]{|#1\rangle}

\begin{document}

\title{Quantum gravitational interaction between a polarizable object and a boundary}
\author{ Jiawei Hu$^1$ and Hongwei Yu$^{2,1,}$\footnote{Corresponding author at hwyu@hunnu.edu.cn}}
\affiliation{$ ^1$ Center for Nonlinear Science and Department of Physics, Ningbo University, Ningbo, Zhejiang 315211, China\\
$^2$ Department of Physics and Synergetic Innovation Center for Quantum Effects and Applications, Hunan Normal University, Changsha, Hunan 410081, China}

\begin{abstract}

We investigate the interaction caused by quantum gravitational vacuum fluctuations between a gravitationally polarizable object and a gravitational boundary, and find a position-dependent energy shift of the object, which induces a force in close analogy to the Casimir-Polder force in the electromagnetic case. For a Dirichlet boundary, the explicit form of the quantum gravitational potential for the polarizable object in its ground-state is worked out and is found to behave like $z^{-5}$ in the near regime, and $z^{-6}$ in the far regime, where $z$ is the distance to the boundary. Taking a Bose-Einstein condensate as a gravitationally polarizable object, we find that the relative correction to the radius caused by fluctuating quantum gravitational waves in vacuum is of order $10^{-21}$. Although far too small to observe in comparison with its electromagnetic
counterpart,  it is nevertheless of the order  of the gravitational strain  caused by a recently detected black hole merger on the arms of the  LIGO.
\end{abstract}

\maketitle

\section{Introduction}Gravitational waves, which are ripples of spacetime, are natural consequences of the theory of general relativity established by Einstein one hundred years ago \cite{einstein1916}. Since their amplitudes  are extraordinarily small, gravitational waves have never been detected directly until the recent breakthrough made by the Laser Interferometer Gravitational-wave Observatory (LIGO) and Virgo collaborations \cite{ligo16} in the culmination of a decades-long quest.  What LIGO  detected are actually  the classical effects of gravitational waves on the length differences between the arms which were revealed by the laser interferometry. Naturally, one may wonder what effects are if gravitational waves are quantized and whether they can be detectable.  Although strictly speaking an ultimate answer of these questions requires a full theory of quantum gravity which is still elusive and quantum gravitational effects are generally expected near the Planck scale, which is far from the energy scales accessible so far, one can still do something using general relativity as  a consistent effective field theory at low energies~\cite{Donoghue94,Hamber95,Khriplovich02,Bjerrum03,Ford15}.  In the present paper,  we are interested in yet another low energy quantum  gravitational  effect, i.e.,  quantum gravitational interaction between a gravitationally polarizable object and a boundary.

A fundamental difference between a quantum theory and a classical one is the quantum fluctuations in vacuum necessitated  by the uncertainty principle. Vacuum fluctuations, although seem fictional, may cause observational effects. A well-known example is the Casimir force between two neutral conducting plates in vacuum \cite{casimir}. Classically, no force other than the universal gravitation would be expected between the two plates because of the absence of external fields. However,  quantum mechanically, the fluctuating quantum  electromagnetic field modes in vacuum are modified due to the presence of reflecting boundaries, and a force is thus induced. One can also replace one of the plates with a neutral polarizable object in the above configuration, and the resulting force is usually referred to as the Casimir-Polder force \cite{cp}. The Casimir and Casimir-Polder effects have played an important role in our understanding of the quantization of electromagnetic fields \cite{casimir2,Bordag09}. One naturally expects that, if gravity has a quantum nature, it should also generate Casimir-like forces.  Here, we plan to calculate the gravitational Casimir-Polder force between a gravitationally polarizable object and a gravitational medium.  Let us note that, the quantum corrections to classical gravitational forces between two polarizable objects from the induced quadrupole moments due to two graviton exchange have recently been studied in Ref.  \cite{Ford15}.


\section{The Basic Formalism}We model a gravitationally polarizable object as a harmonic oscillator. For simplicity, we treat it as a two-level system, which is taken as an open quantum system in interaction with a bath of quantum fluctuating gravitational fields in vacuum. The total Hamiltonian takes the form
\begin{equation}
H=H_A+H_B+H_I\;.
\end{equation}
Here $H_A$ is the Hamiltonian of the two-level system
\begin{equation}
H_A=\hbar\omega_0S_z\;,
\end{equation}
where $S_z={1\/2}\left(\ket{+}\bra{+}-\ket{-}\bra{-}\right)$, and $|+\rangle$ and $|-\rangle$ denote the excited and the ground states, respectively. $H_B$ is the Hamiltonian of the gravitational field, whose explicit expression is not needed here. $H_I$ denotes the quadrupolar gravitational interaction Hamiltonian, which can be written as
\begin{equation}\label{HI1}
H_I=-{1\/2}\sum_{ij}Q_{ij}E_{ij}\;,
\end{equation}
where $Q_{ij}$ is the quadrupole moment of the object and $E_{ij}=-\nabla_i\nabla_j\phi$  with  $\phi$ being the gravitational potential. In Newtonian theory, $E_{ij}$ determines the tidal gravitational acceleration between two nearby test particles, while in general relativity, the similar role is played by the Weyl tensor, i.e. $E_{ij}=-c^2C_{0i0j}$~\footnote{We follow the sign convention of Misner, Thorne, and Wheeler \cite{mtw}.} \cite{Campbell76}. Here $E_{ij}$ and $B_{ij}={1\/2}c^2\epsilon_{imn}C^{mn}_{~~~~0j}$ are the gravito-electric and gravito-magnetic tensors which satisfy the linearized Einstein field equations written in a form in analogy to the Maxwell equations known as gravitoelectromagnetism \cite{matte53,campbell71,szekeres71,maartens98,Ruggiero02,ramos10,Campbell76}.
Note that the gravito-electric field $E_{ij}$ here is supposed to be quantized.

Initially, the whole system is described by $\rho_\text{tot}=\rho(0)\otimes\rho_B$, in which $\rho(0)$ is the initial reduced density matrix of the object, and $\rho_B$ characterizes the state of the environment. The time evolution of the whole system in the frame of the object follows the quantum Liouville equation
\begin{equation}
\frac{\partial\rho_\text{tot}(\tau)}{\partial\tau}=-{i\/\hbar}[H(\tau),\rho_\text{tot}(\tau)]\;.
\end{equation}
The dynamics of the reduced system can be obtained by tracing over the degrees of freedom of the field, and in the limit of weak-coupling, the reduced density matrix is found to satisfy the master equation in the interaction picture \cite{Kossakowski,Lindblad,open}
\begin{equation}\label{master}
{d\/d\tau}\rho(\tau)
=-{i\/\hbar}[H_{LS},\rho(\tau)]+{\cal D}(\rho(\tau))\;.
\end{equation}
We observe from Eq. (\ref{master}) that the contribution of the vacuum gravitational fields can be separated into two parts. The first part
\begin{equation}\label{HLS}
H_{LS}=\hbar\sum_\omega\sum_{ijkl}S_{ijkl}(\omega)A^\dagger_{ij}(\omega){A_{kl}(\omega)}\;,
\end{equation}
is unitary, where $A_{ij}(\omega)=-{1\/2}\sum_{\nu'-\nu=\omega}\Pi(\nu)Q_{ij}\Pi(\nu')$, with $\Pi(\nu)$ denoting the projection onto the eigenspace belonging to the eigenvalue $\nu$ of the Hamiltonian $H_S$. The function $S_{ijkl}(\omega)$ can be written as
\begin{equation}
S_{ijkl}(\omega)= \frac{i}{2}
{\cal{G}}_{ijkl}(\omega)-i\Gamma_{ijkl}(\omega)\;,
\end{equation}
where ${\cal{G}}_{ijkl}(\omega)$ is the Fourier transform of
the field correlation function ${\langle}E_{ij}(s)E_{kl}(0)\rangle$,
\begin{equation}\label{}
{\cal{G}}_{ijkl}(\omega)
={1\/\hbar^2}\int^\infty_{-\infty}dse^{i\omega{s}}{\langle}E_{ij}(s)E_{kl}(0)\rangle\;,
\end{equation}
and $\Gamma_{ijkl}(\omega)$ is the one-side Fourier transform
\begin{equation}
\Gamma_{ijkl}(\omega)={1\/\hbar^2}\int^\infty_0dse^{i\omega{s}}{\langle}E_{ij}(s)E_{kl}(0)\rangle\;.
\end{equation}
Then it can be shown,  
with the help of
\begin{equation}
\frac{1}{x\mp{i}\epsilon}=\text{P}\frac{1}{x}\pm{i}\pi\delta(x)\;,
\end{equation}
that
\begin{equation}\label{}
S_{ijkl}(\omega)=-\frac{\text{P}}{2\pi}\int^\infty_{-\infty}\frac{{\cal{G}}_{ijkl}(\lambda)}{\lambda-\omega}d\lambda\;,
\end{equation}
where P means the principal value. The unitary part $H_{LS}$ is usually referred to as the Lamb shift Hamiltonian, which arises from the object's quadrupolar interaction with  vacuum fluctuations of the gravitational fields and leads to an energy shift  of the object. This part is our main focus in the present Letter, and in the following we will show that this part becomes position-dependent when a boundary is present, which induces a force in close analogy to the Casimir-Polder interaction in the electromagnetic case. The second part
\begin{equation}
{\cal{D}}(\rho)
=\sum_\omega\sum_{ijkl}{\cal{G}}_{ijkl}(\omega)
\bigg(A_{kl}(\omega)\rho A^\dagger_{ij}(\omega)
 -\frac{1}{2}\{A^\dagger_{ij}(\omega)A_{kl}(\omega),\rho\}\bigg)\;,
\end{equation}
is the dissipator of the master equation, which is nonunitary and represents the decoherence and dissipation due to environment.


\section{The Quantum Gravitational Interaction} Now, we calculate the quantum gravitational interaction between a static gravitationally polarizable two-level system whose trajectory can be described as
\begin{equation}
t(\tau)=\tau,~ x(\tau)=y(\tau)=0,~ z(\tau)=z\;,
\end{equation}
and a boundary located at $z=0$. As has been discussed above, one needs to calculate the Lamb shift Hamiltonian (\ref{HLS}) which is related to the correlation functions of the gravito-electric field ${\langle}E_{ij}(s)E_{kl}(0)\rangle$. Consequently, the wave equation in a gravitational medium is needed to fix the boundary conditions, c.f. Ref \cite{Ingraham97}. In the present paper, we apply the Dirichlet boundary condition for simplicity, which models an ideal reflecting boundary for transverse gravito-electric field modes.  
If we expand the spacetime metric as $g_{\mu\nu}=\eta_{\mu\nu}+h_{\mu\nu}$, where $\eta_{\mu\nu}=\{-1,1,1,1\}$ is the Minkowski metric, and work in the transverse traceless (TT) gauge, the gravito-electric field tensor $E_{ij}={1\/2}\ddot{h}_{ij}$, where a dot denotes derivative with respect to $t$. The Wightman function for gravitons in the TT gauge reads \cite{lightcone}
\begin{equation}
\langle h_{ij}(x)\,h_{kl}(x')\rangle
={32\pi G\hbar^2\/c^4}
 (\delta_{ik}\delta_{jl}+\delta_{il}\delta_{jk}-\delta_{ij}\delta_{kl}+D_{ijkl})
 \langle 0|\phi(x)\phi(x')|0 \rangle\;,
\end{equation}
where
\begin{equation}\label{dijkl}
D_{ijkl}=\left(
 {\partial_i\partial'_j\over{\nabla^2}}\delta_{kl}
+{\partial_k\partial'_l\over{\nabla^2}}\delta_{ij}
-{\partial_i\partial'_k\over{\nabla^2}}\delta_{jl}
-{\partial_i\partial'_l\over{\nabla^2}}\delta_{jk}
-{\partial_j\partial'_l\over{\nabla^2}}\delta_{ik}
-{\partial_j\partial'_k\over{\nabla^2}}\delta_{il}
+{\partial_i\partial'_j\partial_k\partial'_l\over\nabla^4}
\right)\;,
\end{equation}
and $\langle 0|\phi(x)\phi(x') |0 \rangle$ is the scalar field two-point function. Here $\nabla^{-2}$ in Eq. (\ref{dijkl}) should be understood in the sense of a Green's function, and when working in momentum space its effect is to bring in a factor of $k^{-2}$. The scalar field two-point function can then be written as the sum of a free space term and a term due to the presence of the boundary with the help of the method of images as
\begin{eqnarray}
\langle 0|\phi(x)\phi(x') |0 \rangle
=&-&{c\/4\pi^2\hbar}\frac{1}{(ct-ct'-i\epsilon)^2-(x-x')^2-(y-y')^2-(z-z')^2}\nn\\
 &+&{c\/4\pi^2\hbar}\frac{1}{(ct-ct'-i\epsilon)^2-(x-x')^2-(y-y')^2-(z+z')^2}\;.
\end{eqnarray}
Since we are interested in the energy shift caused by the boundary, in the following we consider the boundary-dependent terms only.

For a two-level system, the summation over $\omega$ in Eq.~(\ref{HLS}) contains two terms only, i.e. $\omega=\pm\omega_0$. As a result, Eq.~(\ref{HLS}) can be written explicitly as
\begin{eqnarray}\label{HLS2}
H_{LS}&=&\frac{\hbar}{4}\sum_{ijkl}S_{ijkl}(\omega_0)|+\rangle\langle+|Q_{ij}|-\rangle\langle-|Q_{kl}|+\rangle\langle+|\nn
    \\&&+\frac{\hbar}{4}\sum_{ijkl}S_{ijkl}(-\omega_0)|-\rangle\langle-|Q_{ij}|+\rangle\langle+|Q_{kl}|-\rangle\langle-|\;.
\end{eqnarray}
Therefore, the energy-level shifts of the ground state and excited state are
\begin{eqnarray}
&&\delta \mathcal{E}_- =\frac{\hbar}{4}\sum_{ijkl}S_{ijkl}(-\omega_0)\langle-|Q_{ij}|+\rangle\langle+|Q_{kl}|-\rangle\;,\\ &&\delta \mathcal{E}_+
=\frac{\hbar}{4}\sum_{ijkl}S_{ijkl}(\omega_0)\langle+|Q_{ij}|-\rangle\langle-|Q_{kl}|+\rangle\;,
\end{eqnarray}
respectively. Since the external environment is in its vacuum state, we focus on the energy shift of the ground-state, which can be calculated as
\begin{eqnarray}
\delta \mathcal{E}_- ={G\/z^5}\sum_{ijkl}Q_{ij}Q^*_{kl}\,f_{ijkl}(\omega_0,z)\;.
\end{eqnarray}
Here and after we use $Q_{ij}=\langle-|Q_{ij}|+\rangle$, $Q^*_{ij}=\langle+|Q_{ij}|-\rangle$, and $|Q_{ij}|^2=Q_{ij}Q^*_{ij}$ for brevity, and
\begin{equation}
f_{1111}(\omega_0,z)
={\omega_0z\/64\pi c}\int_0^{\infty}du \frac{16u^4+16u^3+20u^2+18u+9}{u^2+\omega_0^2z^2/c^2}e^{-2u}\;,
\end{equation}
\begin{equation}
f_{3333}(\omega_0,z)
={\omega_0z\/8\pi c}\int_0^{\infty}du \frac{4u^2+6u+3}{u^2+\omega_0^2z^2/c^2}e^{-2u}\;,
\end{equation}
\begin{equation}
f_{1122}(\omega_0,z)
=-{\omega_0z\/64\pi c}\int_0^{\infty}du \frac{(2u+1)(8u^3+4u^2-3)}{u^2+\omega_0^2z^2/c^2}e^{-2u}\;,
\end{equation}
\begin{equation}
f_{1133}(\omega_0,z)
=-{\omega_0z\/16\pi c}\int_0^{\infty}du \frac{4u^2+6u+3}{u^2+\omega_0^2z^2/c^2}e^{-2u}\;,
\end{equation}
\begin{equation}
f_{1212}(\omega_0,z)
={\omega_0z\/64\pi c}\int_0^{\infty}du \frac{16u^4+16u^3+12u^2+6u+3}{u^2+\omega_0^2z^2/c^2}e^{-2u}\;,
\end{equation}
\begin{equation}
f_{1313}(\omega_0,z)
=-{\omega_0z\/16\pi c}\int_0^{\infty}du \frac{4u^3+6u^2+6u+3}{u^2+\omega_0^2z^2/c^2}e^{-2u}\;,
\end{equation}
\begin{equation}
f_{1111}(\omega_0,z)=f_{2222}(\omega_0,z)\;,\quad
f_{1122}(\omega_0,z)=f_{2211}(\omega_0,z)\;,\quad
f_{1212}(\omega_0,z)=f_{2121}(\omega_0,z)\;,
\end{equation}
\begin{equation}
f_{1133}(\omega_0,z)=f_{3311}(\omega_0,z)=
f_{2233}(\omega_0,z)=f_{3322}(\omega_0,z)
\end{equation}
\begin{equation}
f_{1313}(\omega_0,z)=f_{3131}(\omega_0,z)
=f_{2323}(\omega_0,z)=f_{3232}(\omega_0,z)\;,
\end{equation}
with other  components being zero.

In analogy to electrodynamics, we define a gravitational polarizability tensor $\alpha_{ij}$ such that $\alpha_{ij}\equiv{|Q_{ij}|^2}/{\hbar\omega_0}$. In the near regime, i.e. when the distance between the object and the reflecting surface is much less than the transition wavelength ($\omega_0z/c\ll1$), the energy shift takes the form
\begin{equation}\label{Eg-near}
\delta \mathcal{E}_- ={3\hbar\omega_0 G\/128z^5}\big(2\alpha_{11}+2\alpha_{22}+17\alpha_{33}+2\alpha_{12}-8\alpha_{13}-8\alpha_{23}\big)\;,
\end{equation}
in which the symmetric and traceless properties of the quadrupole tensor have been taken into account. This shows that the energy shift decays with distance as $z^{-5}$, which can be understood as a gravitational quadrupole-quadrupole interaction between the object and its image. In the long-distance regime, i.e. $\omega_0z/c\gg1$, we have
\begin{equation}
\delta \mathcal{E}_- ={3\hbar Gc\/4\pi z^6}\big(\alpha_{11}+\alpha_{22}+\alpha_{33}+\alpha_{12}-\alpha_{13}-\alpha_{23}\big)\;.
\end{equation}
In this regime, the energy shift decreases with distance as $z^{-6}$, and the factor $c$ appears as a result of retardation.

It has been found in Ref. \cite{Ford15} that the quantum gravitational potential of a couple of polarizable objects are proportional to $z^{-10}$ and $z^{-11}$ respectively in the near and far regimes \cite{Ford15}. The difference in the power law can be understood by dimensional analysis. Here our results are proportional to $\alpha_{ij}$, while those in Ref. \cite{Ford15} are proportional to $\alpha_{ij}^2$. Dimensionally $[\alpha_{ij}]=L^5/G$, then it is reasonable that there is a difference $\propto z^{-5}$ in the dependence of $z$ between the two results.

For a concrete example, we consider a Bose-Einstein condensate (BEC) in a harmonic trap as a gravitationally polarizable object. The BEC will be stretched and squeezed when a gravitational wave passes by, and a gravitational quadrupole will be induced. The gravitational polarizability of the BEC can be calculated in a harmonic oscillator model with the help of the geodesic deviation equation~\cite{szekeres71}, and it is of the order of ${M R^2}/{\omega_0^2}$, where $M$ is the mass of the BEC and $R$ is the radius which can be characterized by the harmonic oscillator length $\sqrt{\hbar/m\omega_0}$ when the interatomic interactions are neglected \cite{BEC}, with $m$ being the mass of a single atom in the BEC, and $\omega_0$ the center-of-mass oscillating frequency which is the same with the trap frequency in the absence of perturbations. The BEC is in interaction with quantum vacuum gravitational fluctuations modified by the presence of a boundary, and a quantum gravitational potential $V_{\rm surf}$ is generated, which will cause a relative shift to the center-of-mass oscillating frequency as \cite{antezza04,Harber05}
\begin{equation}\label{gamma}
\gamma\equiv{\omega_0-\omega\/\omega_0}
\simeq-{1\/2M\omega_0^2}\frac{\partial^2}{\partial z^2}V_{\rm surf}\;.
\end{equation}
If the frequency of a harmonic oscillator is suddenly changed from $\omega_0$ to $\omega$ at some position $z=z_0$, its amplitude will be changed from $R$ to $R'$, and the variables can be related to each other on equalling the kinetic energy of the oscillator at $z=z_0$ as \cite{szekeres71}
\begin{equation}\label{R}
R'^2=R^2+\frac{\omega_0^2-\omega^2}{\omega^2}(R^2-z_0^2)\;.
\end{equation}
Plugging Eq. (\ref{gamma}) and $\langle z_0^2\rangle=R^2/2$ into Eq. (\ref{R}), we have
\begin{equation}
{R'-R\/R}\simeq {1\/2}\gamma\;.
\end{equation}
That is, due to the presence of $V_{\rm surf}$, both the oscillating frequency and the radius of the BEC are modified, which are both of the order of $\gamma$. For a BEC composed of $N=10^6$ $^{87}$Rb atoms trapped with $\omega_0\sim 10^2~{\rm Hz}$, the typical size $R\sim 1~{\rm \mu m}$. We assume that the center of masss of the BEC is located at a distance $z\simeq R\sim  1~{\rm \mu m}$ to the boundary, then $\gamma\sim\frac{\hbar G R^2}{z^7\omega_0^3}\sim 10^{-21}$. Here let us note that the relative oscillating frequency shift of a trapped BEC has been utilized to detect the electromagnetic Casimir-Polder force, where  $\gamma\sim 10^{-2}-10^{-4}$ in the retarded and thermal regimes \cite{antezza04,Harber05}. In the electromagnetic case, $V_{\rm surf}$ is calculated as a summation of the Casimir-Polder potentials of individual atoms, while in the gravitational case here, it is calculated by considering the BEC as a whole system.  The relative correction  is exceedingly small  as compared to that in the electromagnetic case, so,  far too small to observe. It is interesting to  note however that the  relative correction of the radius of the BEC $\gamma\sim 10^{-21}$ caused by the quantum gravitational interaction is of the same order of the gravitational strain  caused by a black hole merger on the arms of the  LIGO~\cite{ligo16}. Noteworthily, the relative correction here is caused by fluctuating quantum  gravitational waves in vacuum whereas the  correction  observed by LIGO was caused by classical gravitational waves produced by  a binary black hole merger.
We must point out that an experimental verification of this quantum gravitational effect would be a much greater challenge than that of  the classical one detected by LIGO, even if we could find matter that would reflect gravitational waves significantly, since the radius of the BEC is a lot more difficult to be measured precisely. 

Now a few comments are in order for our assumption of a
plane that perfectly reflects gravitational waves.  First, let us note that the propagation of gravitational waves in material media was studied in Ref. \cite{Ingraham97}, and   the reflection coefficients for gravitational waves at the interface of two gravitational media is derived, which is a function of the gravitational susceptibility $\chi$, or equivalently the refractive index $n$. Microscopically, $n$ is related to the gravitational polarizability of the molecules the medium is composed of, and is found to be \cite{szekeres71}
\begin{equation}
n\simeq 1+\frac{3}{8\theta g^2}A^3/D^3,
\end{equation}
where $A$ is the average linear dimension of a typical molecule, $D$ is the mean distance between molecules, $g=\lambda/2\pi A$ and $\theta=\omega_0^2 A^3/mG$ are dimensionless variables, with $\lambda$ being the wavelength, $m$ and $\omega_0$ the mass and characteristic frequency of the molecules modelled as harmonic oscillators. Here ``molecule" refers to any basic unit that constitutes the medium and therefore is a term in a general sense. For ordinary materials which are bounded electrically, $\theta\sim 10^{40}$ \cite{szekeres71}. That is, ordinary materials can hardly be polarizable by gravitational waves, and the reflection coefficient for gravitational waves will be extremely small. However, if the molecules of the materials are bounded gravitationally, $\theta\sim1$. For example,  for a medium whose molecules are stars with an appropriate internal equation of state, it is plausible  that $\theta<1$ \cite{szekeres71}. Furthermore, if the medium behaves like a solid or liquid $(A/D\sim1)$, the refractive index may deviate from unity considerably, and the reflection may be significant when the wavelengths is not much larger than the typical length of the molecules, i.e., $g\sim1$. Therefore,  if there were such media in our Universe, then the propagation of certain gravitational waves through them might be noticeably affected. Second, there have  been interesting suggestions that the interaction between gravitational waves and quantum fluids (e.g. superconductors, superfluids, quantum Hall fluids, and Bose-Einstein condensates) might be enhanced compared with ordinary matter (See Ref. \cite{Kiefer} for a review), and even some work has been done on possible implications of these suggestions~\cite{gc}. Finally, it is interesting to revisit the quantum effects we just  studied using a more realistic model of a gravitational boundary rather than a perfect reflector, which, though a much more complicated issue, is currently under investigation.



\section{Conclusion} In summary, we have studied the interaction caused by fluctuating quantum vacuum gravitational fields between a polarizable object modelled as a gravitational two-level system and a boundary. 
 The position-dependent potential induces a force in close analogy to the Casimir-Polder force. We have worked out the explicit analytical expressions of the quantum gravitational interaction potential for a Dirichlet boundary, which decreases with distance as $z^{-5}$ in the short-distance regime, and $z^{-6}$ in the long-distance regime. Taking a Bose-Einstein condensate as a gravitationally polarizable object,, we have found that the relative correction to the radius caused by fluctuating quantum  gravitational waves in vacuum  is of the  order $10^{-21}.$


{\it Acknowledgments}.---We would like to thank Wenting Zhou for very helpful discussions. This work was supported in part by the NSFC under Grants No. 11375092, No. 11435006, and No. 11447022;  the Zhejiang Provincial Natural Science Foundation of China under Grant No. LQ15A050001; the Research Program of Ningbo University under Grant No. XYL15020 and No. xkzwl1501; and the K. C. Wong Magna Fund in Ningbo University.


\begin{thebibliography}{00}

\bibitem{einstein1916}
A. Einstein, Sitzungsber. K. Preuss. Akad. Wiss. {\bf 1}, 688 (1916).

\bibitem{ligo16}
B. P. Abbott {\it et al.}, Phys. Rev. Lett. {\bf 116}, 061102 (2016).

\bibitem{Donoghue94}
J. F. Donoghue, Phys. Rev. Lett. {\bf 72}, 2996 (1994); Phys. Rev. D {\bf 50}, 3874 (1994).

\bibitem{Hamber95}
H. W. Hamber and S. Liu, Phys. Lett. B {\bf 357}, 51 (1995).

\bibitem{Khriplovich02} 
I. B. Khriplovich and G. G. Kirilin, J. Exp. Theor. Phys. {\bf 95}, 981 (2002) [Zh. Eksp. Teor. Fiz. {\bf 95}, 1139 (2002)].

\bibitem{Bjerrum03} 
N. E. J. Bjerrum-Bohr, J. F. Donoghue and B. R. Holstein, Phys. Rev.
D {\bf 67}, 084033 (2003); Erratum: {\bf 71}, 069903 (2005).

\bibitem{Ford15}
L. H. Ford, M. P. Hertzberg, and J. Karouby, Phys. Rev. Lett. {\bf 116}, 151301 (2016) [arXiv:1512.07632 [hep-th]].




\bibitem{casimir}
H. B. G. Casimir, Proc. K. Ned. Akad. Wet. {\bf 51}, 793 (1948).

\bibitem{cp}
H. B. G. Casimir and D. Polder, Phys. Rev. {\bf 73}, 360 (1948).

\bibitem{casimir2}
G. L. Klimchitskaya, U. Mohideen, and V. M. Mostepanenko, Rev. Mod. Phys. {\bf 81}, 1827 (2009).

\bibitem{Bordag09}
M. Bordag, G. L. Klimchitskaya, U. Mohideen, and V. M. Mostepanenko, {\it Advances in the Casimir Effect} (Oxford University Press, Oxford, 2009).



\bibitem{mtw}
C. W. Misner, K. S. Thorne, and J. A. Wheeler, {\it Gravitation} (W. H. Freeman and Company, San Francisco, 1973).

\bibitem{Campbell76}
W. B. Campbell and T. A. Morgan, Am. J. Phys. {\bf 44}, 356 (1976).

\bibitem{matte53}
A. Matte, Can. J. Math. {\bf 5}, 1 (1953).

\bibitem{campbell71}
W. Campbell and T. Morgan, Physica (Amsterdam) {\bf 53}, 264 (1971).

\bibitem{szekeres71}
P. Szekeres, Ann. Phys. (N.Y.) {\bf 64}, 599 (1971).

\bibitem{maartens98}
R. Maartens and B. A. Bassett, Classical Quant. Grav. {\bf 15}, 705 (1998).

\bibitem{Ruggiero02}
M. L. Ruggiero and A. Tartaglia, Nuovo Cimento B {\bf 117}, 743 (2002).

\bibitem{ramos10}
J. Ramos, M. de Montigny, and F. Khanna, Gen. Relativ. Gravit. {\bf 42}, 2403 (2010).



\bibitem{Kossakowski}
V. Gorini, A. Kossakowski, and E. C. G. Surdarshan, J. Math. Phys. {\bf 17}, 821 (1976).

\bibitem{Lindblad}
G. Lindblad, Commun. Math. Phys. {\bf 48}, 119 (1976).

\bibitem{open}
H.-P. Breuer and F. Petruccione, {\it The Theory of Open Quantum Systems} (Oxford University Press, Oxford, 2002).


\bibitem{Ingraham97}
R. Ingraham, Gen. Relativ. Gravit. {\bf 29}, 117 (1997).

\bibitem{lightcone}
H. Yu and L. H. Ford, Phys. Rev D {\bf 60}, 084023 (1999).



\bibitem{BEC}
F. Dalfovo, S. Giorgini, L. P. Pitaevskii, and S. Stringari, Rev. Mod. Phys. {\bf 71}, 463 (1999).

\bibitem{antezza04}
M. Antezza, L. P. Pitaevskii, and S. Stringari, Phys. Rev. A {\bf 70}, 053619 (2004).

\bibitem{Harber05}
D. M. Harber, J. M. Obrecht, J. M. McGuirk, and E. A. Cornell, Phys. Rev. A {\bf 72}, 033610 (2005).

\bibitem{Kiefer}
C. Kiefer and C. Weber, Ann. Phys. (Leipzig) {\bf 14}, 253 (2005).

\bibitem{gc}
J. Q. Quach, Phys. Rev. Lett. {\bf 114}, 081104 (2015).



\end{thebibliography}
\end{document}